\documentclass{article}

\usepackage{PRIMEarxiv}

\usepackage[utf8]{inputenc} 
\usepackage[T1]{fontenc}    
\usepackage{mathtools} 
\usepackage{amsthm, amsfonts, amssymb} 
\usepackage{hyperref}       
\usepackage{url}            
\usepackage{booktabs}       
\usepackage{amsfonts}       
\usepackage{nicefrac}       
\usepackage{microtype}      
\usepackage{lipsum}
\usepackage{fancyhdr}       
\usepackage{graphicx}       
\graphicspath{{media/}}     
\usepackage{siunitx}

\usepackage{graphicx}
\usepackage[dvipsnames]{xcolor}
\usepackage{setspace, caption, tikz, mdframed}
\usepackage[position=b]{subcaption} 
\usepackage{svg}
\usepackage{todonotes}

\title{Improved Anisotropic Gaussian Filters
}

\author{
  Alex Keilmann \\
  University of Kaiserslautern-Landau \\
  \& Fraunhofer ITWM\\
  Gottlieb-Daimler-Str. 49, 67663 Kaiserslautern\\
  \texttt{keilmann@rptu.de} \\
   \And
  Michael Godehardt \\
  Image Processing Department\\
  Fraunhofer ITWM \\
  Fraunhofer Platz 1, 67663 Kaiserslautern\\
  \And
  Ali Moghiseh \\
  Image Processing Department\\
  Fraunhofer ITWM \\
  Fraunhofer Platz 1, 67663 Kaiserslautern\\
   \AND
   Claudia Redenbach \\
   Mathematics Department\\
   University of Kaiserslautern-Landau \\
   Gottlieb-Daimler-Str. 49, 67663 Kaiserslautern \\
   \And
   Katja Schladitz \\
   Image Processing Department\\
   Fraunhofer ITWM \\
   Fraunhofer Platz 1, 67663 Kaiserslautern \\
}

\begin{document}
\maketitle

\begin{abstract}
Elongated anisotropic Gaussian filters are used for the orientation estimation of fibers. In cases where computed tomography images are noisy, roughly resolved, and of low contrast, they are the method of choice even if being efficient only in virtual 2D slices. However, minor inaccuracies in the anisotropic Gaussian filters can carry over to the orientation estimation. 
Therefore, this paper proposes a modified algorithm for 2D anisotropic Gaussian filters and shows that this improves their precision. Applied to synthetic images of fiber bundles, it is more accurate and robust to noise. Finally, the effectiveness of the approach is shown by applying it to real-world images of sheet molding compounds.
\end{abstract}

\keywords{Computed tomography \and Directional filter \and Fiber direction \and Fiber reinforced polymers \and Orientation estimation \and Sheet molding compounds}

\section{Introduction}
\label{sec1}

Gaussian filters have a wide variety of applications in image processing. Whereas isotropic Gaussian filters, being the foundation of scale space theory \cite{lindeberg96}, can be implemented easily, their anisotropic counterparts are more demanding while being just as interesting \cite{lampert06}: They give a handle on orientation as well as scale, which makes them the cornerstones of orientation space theory \cite{faas03}.

Anisotropic Gaussian filters have been employed for denoising images \cite{yang96, treece20} as they can be adapted to image structures and, hence, preserve edges. Another application is the estimation of orientations using a filter bank of anisotropic Gaussian filters. For example, local fiber directions can be estimated by the direction of the maximal response of anisotropic Gaussian filters that are aligned along a given set of directions \cite{robb07, wirjadi09a}.

{Most established methods for estimating fiber directions are based on gradients, such as calculating the image's Hessian matrix \cite{eberly94,Frangi98multiscalevessel,ohser-schladitz09book} or the second-moment matrix of the image's gradient, called the \textit{structure tensor} \cite{haglund92, weickert99,krause10}. In both methods, the local fiber directions result from the eigendecomposition of the respective matrix. Wirjadi et~al.~ \cite{wirjadi16} compared them to other methods based on 3D images of single synthetic fibers and identified them as the most accurate.  Pinter et~al.~
\cite{pinter18} further investigated their accuracy on images of multiple fibers, finding the structure tensor to be comparably more robust.  

In their comparison,} Wirjadi et~al.~\cite{wirjadi16} and Pinter et~al.~\cite{pinter18} both identified the Maximal Response (\textit{MR}) method as robust to noise but suffering from the trade-off between runtime and accuracy in 3D. 
However, there are {use} cases where the image quality 
{is too low for} methods based on local gray-value derivatives on the one hand and where due to the production process the fibers are known to be oriented in a 2D subspace anyway. This holds for instance true for \textmu CT images of sheet molding compounds, {a material where reinforcing fibers} lie within a plane. Then, only 2D images have to be analyzed{, similarly to stereological approaches}. In {2D}, the MR method is less restricted regarding runtime and even outperforms other methods due to its robustness with respect to low image contrast \cite{schladitz16}. 

The accuracy of the direction estimation clearly depends on the accuracy of the filter responses for the considered directions. Under otherwise perfect conditions, this method's results barely depend on contrast as the filter responses scale with the contrast. Although the response differences are smaller, this does not influence which response is maximal. However, computed tomography images are often affected by noise and other artifacts. For low-contrast images, these effects have a much stronger impact on the detected direction of maximal response due to the small differences in responses for varying angles. In this case, small inaccuracies in the anisotropic Gaussian filter can impair the direction estimation further. In this paper, we will consider the case of low contrast between the foreground, i.e., fibers, and noise, while using a low resolution for the fibers.

Anisotropic Gaussian filters in $\mathbb{R}^2$ can be implemented naïvely by filtering in the directions of the major and the minor axis of the  Gaussian's contour ellipses, subsequently. However, Geusebroek et~al.~\cite{geusebroek03} derived a more accurate decomposition, where at least one of two filter directions is aligned with an axis of the image grid. Whereas the naïve implementation may need interpolation for filtering in both directions, Geusebroek \textit{et~al.}'s method requires interpolation for at most one filter direction. Lampert\&Wirjadi~\cite{lampert06} generalized these results to $\mathbb{R}^d$ and provided explicit formulas for $\mathbb{R}^3$.

Besides implementational inaccuracies of the 1D filters, interpolation introduces spatial inhomogeneity into the filter kernels, as Lam \& Shi~\cite{lam2007recursive} have shown. Therefore, they propose a modification of Geusebroek \textit{et~al.}'s method which avoids interpolation altogether at the cost of an additional 1D Gaussian filtering step. However, Lam and Shi's modification limits possible half-axis ratios 
$\omega = {\sigma_2}/{\sigma_1}$, 
to $\omega \geq 0.4142$. The ratio can be lowered to $\omega \geq 0.1622$ at the cost of aliasing effects. In our setting, far smaller ratios are needed to accurately mimic the fiber shape, e.g., $\omega = 0.025$ in {the Application section.}

In this paper, we therefore suggest another modification not suffering from this restriction. 
{W}e propose a modification to Geusebroek's decomposition that halves the number of interpolation steps 
{and} show that this modification improves performance. Moreover, we consider cubic instead of linear interpolation, which improves accuracy at the cost of speed. Based on synthetic fiber images, we show that the adapted method results in higher accuracy of the maximal response method. {Additionally, we show that it outperforms estimation based on the Hessian matrix or the structure tensor in 2D.} Finally, we apply our method to real-world images of sheet molding compounds{.}

\section{Materials and Methods}
\label{sec:MaxResponse}
A natural approach to calculating the anisotropic Gaussian filter in $\mathbb{R}^ d$ is to decompose it into a sequence of multiple Gaussian filters in $\mathbb{R}$, which poses a simpler problem \cite{lampert06}. The recursive scheme with infinite impulse response by Young \textit{et~al.}~\cite{young95, young02}, using boundary conditions by Triggs \& Sdika~\cite{triggs06} has proven efficient and accurate.
For the case of $\mathbb{R}^2$, Geusebroek et~al.~\cite{geusebroek03} propose a decomposition into filters along the $x_1$-axis of the image grid and a filter along another direction that generally does not align with the grid.

Initially, consider an axis-aligned Gaussian kernel with standard deviations $\sigma_1 > \sigma_2 > 0$ centered in the origin, i.e.,
\begin{align}
    \label{eq:aniGauss}
    g_{\sigma_1, \sigma_2}(x_1, x_2) = &\frac{1}{\sqrt{2 \pi} \sigma_1} \exp\left(-\frac{1}{2} \frac{x_1^2}{\sigma_1^2}\right)\cdot \\&\frac{1}{\sqrt{2 \pi} \sigma_2} \exp\left(-\frac{1}{2} \frac{x_2^2}{\sigma_2^2}\right), ~~~ x_1, x_2 \in \mathbb{R}.
\end{align}
Its contour lines are axis-aligned ellipses with half-axis ratio $\omega= \sigma_2/\sigma_1$. We now rotate the kernel to get
$g_{\sigma_1, \sigma_2, \theta}$, whose major half axis points in direction $\bf{\nu}=(\cos(\theta), \sin(\theta))^T$ for $\theta \in [0, \pi)$. Formally,
\begin{align}
\label{eq:aniGaussMainAxisDecomp}
    g_{\sigma_1, \sigma_2, \theta}(x_1, x_2)
    = &\frac{1}{\sqrt{2 \pi} \sigma_1} \exp\left(-\frac{1}{2} \frac{(\pmb{x}^T \pmb{\nu})^2}{\sigma_1^2}\right) \\&\frac{1}{\sqrt{2 \pi} \sigma_2} \exp\left(-\frac{1}{2} \frac{(\pmb{x}^T\pmb{\nu^{\bot}})^2}{\sigma_2^2}\right),
\end{align}
where $\pmb{x}=(x_1, x_2) \in \mathbb{R}^2$ and $\pmb{\nu^{\bot}}=(-\sin(\theta), \cos(\theta))^T.$

A decomposition of the corresponding filter into one-dimensional filters along the coordinate axes is generally not possible. However, Geusebroek et~al.~\cite{geusebroek03} proved that a decomposition into filters along the $x_1$-direction and the direction 

\begin{flalign}
    \nu_\ast &= \nu_\ast(x_1, x_2, \theta, \sigma_1, \sigma_2) = x_1 \cos (\varphi) + x_2 \sin (\varphi) \\
    &\text{~with} \nonumber\\
    \tan \varphi &= \frac{\sigma_2^2\cos^2 \theta + \sigma_1^2\sin^2 \theta}{\left(\sigma_1^2 - \sigma_2^2 \right) \cos \theta \sin \theta}.
\end{flalign}
is indeed possible, namely with the kernel
\begin{align}
    \label{eq:aniGaussCoordAxisDecomp}
    g_{\sigma_1, \sigma_2, \theta}(x_1, x_2) = &\frac{1}{\sqrt{2 \pi} \sigma_x} \exp\left(-\frac{1}{2} \frac{x_1^2}{\sigma_x^2}\right) 
    \\&
    \frac{1}{\sqrt{2 \pi} \sigma_{\nu_\ast}} \exp\left(-\frac{1}{2} \frac{\nu_\ast^2}{\sigma_{\nu_\ast}^2}\right),
     ~~~ x_1, x_2 \in \mathbb{R}.
\end{align}
The standard deviations $\sigma_x, \sigma_{\nu_\ast}$ can be computed in terms of the rotation angle $\theta$ and the standard deviations $\sigma_1, \sigma_2$ via 
\begin{align}
    \sigma_x &= \sigma_x(\theta, \sigma_1, \sigma_2) ~= \frac{\sigma_1 \sigma_2}{\sqrt{\sigma_1^2\cos^2 \theta + \sigma_2^2\sin^2 \theta}}\\
    \sigma_{\nu_\ast} &= \sigma_{\nu_\ast}(\theta, \sigma_1, \sigma_2) = \frac{1}{\sin \varphi} \sqrt{\sigma_2^2\cos^2 \theta + \sigma_1^2\sin^2 \theta} 
\end{align}

\begin{figure}[!t]
\centering
\begin{subfigure}{.49\textwidth}
    \centering
    \begin{tikzpicture}[scale=1]
        \draw [dashed, gray, ->](-2.2,0) -- (2.2,0) node [above] { $x_1$};
        \draw [dashed, gray, ->] (0,-2.2) -- (0,2.2) node [above] {$x_2$};
        \draw [very thick, ->] (-2, -2) -- (2, 2) node [below] { $~~~v_1$};
        \draw [very thick, ->] (2, -2) -- (-2, 2) node [below] {$v_2~~$};
        \draw[ultra thick, rotate=45] (0,0) ellipse (2cm and 1cm);
        
        
        \draw [thick] (6mm,0pt) arc(0:45:6mm);
        \draw [ultra thick](3.5mm, 1.5mm) node{$\theta$};
        
    \end{tikzpicture}
    \caption{}
\end{subfigure}
\begin{subfigure}{.49\textwidth}
\centering
    \begin{tikzpicture}[scale=1]
        \draw [very thick, ->](-2.2,0) -- (2.2,0) node [above] {$x_1$};
        \draw [dashed, gray, ->] (0,-2.2) -- (0,2.2) node [above] { $x_2$};
        \draw [gray, ->] (-2, -2) -- (2, 2) node [below] {$~~~~~v_1$};
        \draw [gray, ->] (2, -2) -- (-2, 2) node [below] {$v_2~~$};
        \draw [very thick , ->](-1.1, -2.2) -- (1.1, 2.2) node [below] {$~~~~\nu_\ast$};
        \draw[ultra thick, rotate=45] (0,0) ellipse (2cm and 1cm);
    	\draw  [very thick](6mm,0pt) arc(0:63:6mm);
        \draw [very thick](3.4mm, 1.9mm) node{$\varphi$};
    \end{tikzpicture}
    \caption{}
\end{subfigure}
\caption{The Gaussian ellipse, i.e.,~contour line of the Gaussian function, w.r.t. (a) the principal axes $v_1$ and $v_2$, and (b) the axes $x_1$ and $\nu_\ast$ \cite{geusebroek03}.}
\label{fig:geusebroekShearing}
\end{figure}
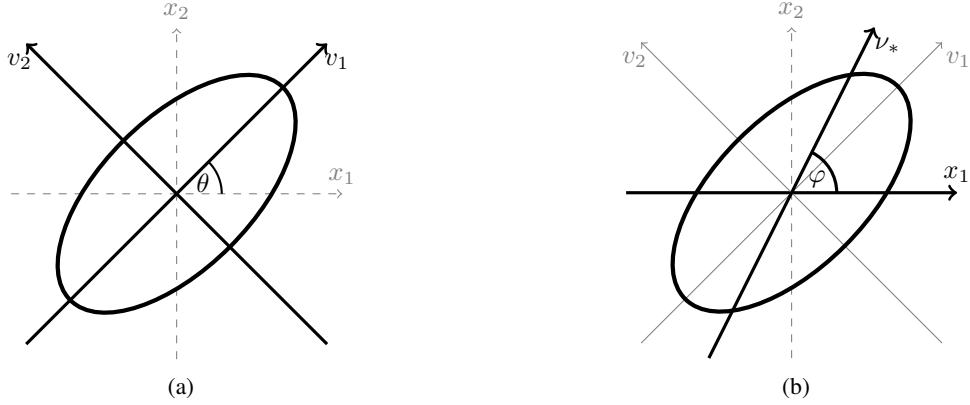

Fig.~\ref{fig:geusebroekShearing} illustrates this decomposition; see Geusebroek et~al.~\cite{geusebroek03} for the detailed derivation.

\subsection{Algorithms for Anisotropic Gaussian Filtering}
Gaussian filters can be implemented naïvely by rotating the image with the same matrix that rotates the filter such that its major and minor axes are aligned with the coordinate axes. Then, the image can be filtered along the coordinate axes using Young \textit{et~al.}'s~\cite{young02} recursive Gaussian filter. {This corresponds to the filter decomposition along the principal axes, see Eq.~\ref{eq:aniGaussMainAxisDecomp}.}
This way, more memory is consumed as the image does not fit its previous rectangular structure anymore. Moreover, interpolation steps are necessary for both filter directions~\cite{lampert06}.

{More advanced algorithms make use of decomposition along other axes, as in Eq.~\ref{eq:aniGaussCoordAxisDecomp}:} 
Lampert \& Wirjadi's~\cite{lampert06} \textit{geometric algorithm} circumvents the interpolation along one axis by considering the filter decomposition as a shear of the coordinate axes with a shear matrix $V$. Hence, the image is sheared with $V$ before filtering along the coordinate axes. Afterwards, the resulting image is transformed back with $V^{-1}$.

In Geusebroek \textit{et~al.}'s~\cite{geusebroek03} \textit{line buffer algorithm}~\cite{lampert06} the image is processed in-place as it filters along the $x_1$-axis and the $\nu_\ast$-line, see above. The transformation step necessary for filtering along the $\nu_\ast$-line is the inverse shear used in the geometric algorithm. This transformation using interpolation is necessary every time data is read or written. This can be kept minimal by using image line buffers for the filter history. However, as the recursive Gaussian filter consists of a forward and a backward filter, this yields 2 forward and 2 backward transformation steps, yielding 4 interpolation steps per pixel. 



In comparison, the geometric algorithm uses only 2 transformations and, thus, interpolations per pixel, which makes it less error-prone compared to the line buffer algorithm. However, the geometric algorithm needs more memory because the transformed image no longer fits the original rectangular shape.

\subsection{The Hybrid Algorithm}
\label{sec:hybrid}
Our improved scheme combines the advantages of both the geometric and the line buffer algorithm:
It filters in $x_1$-direction with Young et~al.'s~\cite{young02} recursive Gaussian filter as in the line buffer algorithm. The filter in $\nu_\ast$-direction is modified such that the intermediate transformation steps are omitted: As the forward and backward filter move along the same line, the intermediate transformation steps taken together are the identity. Therefore, the result of the forward filter does not need to be transformed but can be stored in-place. This approach requires 2 interpolation steps per pixel, as in the geometric algorithm, while using as little memory as the line buffer algorithm. The difference to the established algorithms is  "smarter bookkeeping".
Hereafter, we will call this the \textit{hybrid algorithm}\footnote{{A free, open-source implementation in C++ is available at \hyperlink{https://github.com/akeilmann/aniGauss}{https://github.com/akeilmann/aniGauss}.}}.

An axis-aligned filter is generally more accurate than a filter that is not axis-aligned since the latter requires interpolation. Therefore, we first filter along the axis and, subsequently, in $\nu_\ast$-direction.

So far, we only discussed a decomposition into filters where one filter direction is aligned with the $x_1$-axis. Analogously, a decomposition such that one filter direction is aligned with the $x_2$-axis is possible \cite{lampert06}. This may even be advantageous for $45^\circ \leq \theta \leq 135^\circ$: The standard deviation $\sigma_x$ of the filter along the $x_1$-axis varies over the rotation angles $\theta$, being largest for $\theta = 0^\circ$ and smallest for $\theta = 90^\circ$. For the line buffer and the hybrid algorithm, filtering along the $x_1$-axis smoothes the image in the same direction, in which the interpolation takes place. This may be less error-prone for stronger smoothing. Hence, we propose to decompose the anisotropic filter with an $x_2$-aligned axis for $45^\circ \leq \theta \leq 135^\circ$.

This modification is possible for each of the approaches mentioned above. In the following, we call this the \textit{major-axis modification}.

\subsection{Theoretical Performance Analysis}
The runtime of the anisotropic Gaussian filter is constant for each pixel and depends only on the rotation angle and not on the variance.
The filtering steps require $12$ additions and $13$ multiplications. 
Linear interpolation can be implemented with $1$ addition and $2$ multiplications per pixel.

We further propose to apply cubic interpolation with natural boundary conditions. In our implementation, each cubic interpolation step takes $8$ additions and $14$ multiplications per pixel. Therefore, we only combine it with the hybrid algorithm. The total complexities per pixel are listed in Table~\ref{tab:complexity}.
\begin{table}[!t]
    \begin{center}
        \caption{Complexity per pixel for different algorithms with interpolation. \label{tab:complexity}}
        \begin{tabular}{|l|c|c|c|}
            \hline
            & Line buffer & \multicolumn{2}{c|}{Hybrid}\\
            &Linear&Linear&Cubic\\
            \hline
            Multiplications & 21 & 17 & 27\\
            Additions & 16 & 14 & 20 \\
            \hline
        \end{tabular}
    \end{center}
\end{table}

The runtime of the MR method in total depends on the complexity of the employed anisotropic filter algorithm, the discretization of the direction space, i.e., the number of angles considered, and the image size. The dependency on the latter three is linear, thus we discuss the speed of Gaussian filters, only.

\subsection{{Fiber Orientation Estimation}}
\label{sec:fiberOrientationEstimation}
\subsubsection{{Maximal Response of Anisotropic Gaussian Filters}}
To filter an image of fibers for directions, imitate the elongated shape of a fiber
with the $d$-dimensional \emph{anisotropic Gaussian (function)} $g_{\theta}$, see Eq.\,\ref{eq:aniGauss}.
Its parameters give a handle on the orientation, length, and diameter for the fiber model \cite{lampert06}.

The filter response $(\pmb{g}_{\theta} \ast \pmb{f})(\pmb{x} )$ to the image $\pmb{f}$ is maximal when $\theta$ matches the local fiber direction in the point $\pmb{x}  \in \Xi$, where $\Xi$ is the fiber system. Therefore, one can find the direction $\nu$ that maximizes the filter response for all $\pmb{x} \in \Xi$ \cite{wirjadi16}:
\begin{equation}
    \nu(\pmb{x} ) = \underset{\theta \in {S^{1}_+}}{\operatorname{argmax}}(\pmb{g}_{\theta} \ast \pmb{f})(\pmb{x} )
\end{equation}

$\nu(\pmb{x} )$ is estimated by calculating the convolution$(\pmb{g}_{\theta} \ast \pmb{f})(\pmb{x} )$ for a finite set of directions that covers the space as evenly as possible \cite{schladitz16}.
Hereafter, we will call this the \textit{MR method}.

{
The method's accuracy is mainly influenced by the parameter $\sigma_2$, which we will set to ${r}/{2}$ in the following.
This is motivated by the $2 \sigma$ rule for the normal distribution, which says that approximately $95\%$ of the data points are within two standard deviations of the mean \cite{Georgii2012}. Hence, a correctly aligned filter kernel $g_\theta$ that is centered within the fiber covers the fiber's thickness with $95\%$ of its weight when $\sigma_2 = {r}/{2}$, where $r$ is the fiber's radius. Pixels that are further away than $2\sigma_2$ are barely taken into account. This ensures that the filter response is maximal when the 2-dimensional Gaussian filter kernel is aligned with the fiber:
a much larger variance might take too many pixels outside of the fiber into account, while a much smaller variance results in filter kernels whose main mass is concentrated in an elliptical region that is thinner than the fibers. In this case, the ellipse might fit inside the fiber for several angles $\theta$ which makes it harder to accurately detect the direction for which the maximum is attained.
The parameter $\sigma_1$ corresponds to the elongation of the fiber. We observed that it has rather little influence on the estimation accuracy. Yet, it should hold $\sigma_1< {\ell}/{4}$, where $\ell$ is the fiber length, analogous to the argument above, and the Gaussian kernel should still be clearly elongated.}

\subsubsection{{Structure Tensor}}
{
The image's gradient describes directions in the image as it is aligned with the fiber's surface normal, which in turn is orthogonal to the fiber direction. Further convolution with a Gaussian kernel allows for robust direction estimation against noise. 

For smoothing parameters $\sigma, \rho > 0$, let the isotropic Gaussian kernel $g_\sigma := g_{\sigma, \sigma}$ and $\nabla f$ the gradient of the image $f$. The structure tensor is defined as 
\begin{equation}
    S_{\sigma, \rho} = g_\rho \ast \left(\nabla (f\ast g_\sigma) \cdot \nabla (f\ast g_\sigma)^T \right)
\end{equation}
as denoted in Wirjadi et~al.~\cite{wirjadi16} and proposed by Haglund~\cite{haglund92}. Its eigenvector corresponding to the smallest eigenvalue describes the local fiber direction.

As argued and done by Krause et~al.~\cite{krause10} and Wirjadi et~al.~\cite{wirjadi16}, we set $\sigma = r$. This choice was empirically confirmed by Pinter et~al.~\cite{pinter18}. The parameter $\rho$ is less impacted by the size of the underlying structure. Hence, we use a constant $\rho = 6.0$, as was found optimal by Pinter et~al.~\cite{pinter18}.}

\subsubsection{{Hessian Matrix}}
{
The image's Hessian matrix
\begin{equation}
    H_\sigma = \nabla \nabla^T (f\ast g_\sigma)
\end{equation}
with smoothing parameter $\sigma > 0$ describes image curvature, which is minimal along the fiber direction. Therefore, the local fiber direction is determined by the Hessian's eigenvector corresponding to the smallest eigenvalue. Analogous to the structure tensor, we set $\sigma = r$ following Wirjadi et~al.~\cite{wirjadi16} and Pinter et~al.~\cite{pinter18}.}

\section{Results}
\label{sec:Experiments}
In this section, we support the theoretical analysis with experimental results. More precisely, we show that the hybrid algorithm is more accurate than the line buffer algorithm. Employing linear interpolation, the hybrid algorithm is indeed faster than the line buffer algorithm. In the first subsection, we test the performance of anisotropic Gaussian filters as such. In the second subsection, we test performance on synthetic fiber bundles with varying noise contrast to the background. In the third subsection, we apply the algorithms to real-life data.

The following experiments were carried out on an Intel(R) Core(TM) i7-7500U CPU @2.70\,GHz with 16\,GiB of RAM, using the GNU compiler GCC\,9.0 on a 64-bit GNU/Linux operating system.

\subsection{Performance of Anisotropic Gaussian Filters}
We test the performance of the anisotropic Gaussian filters for the line buffer algorithm using linear interpolation and for the hybrid algorithm using linear interpolation as well as cubic interpolation with natural boundary conditions.

\begin{figure*}[!t]%
\begin{subfigure}{\textwidth}
    \includegraphics[width=\textwidth]{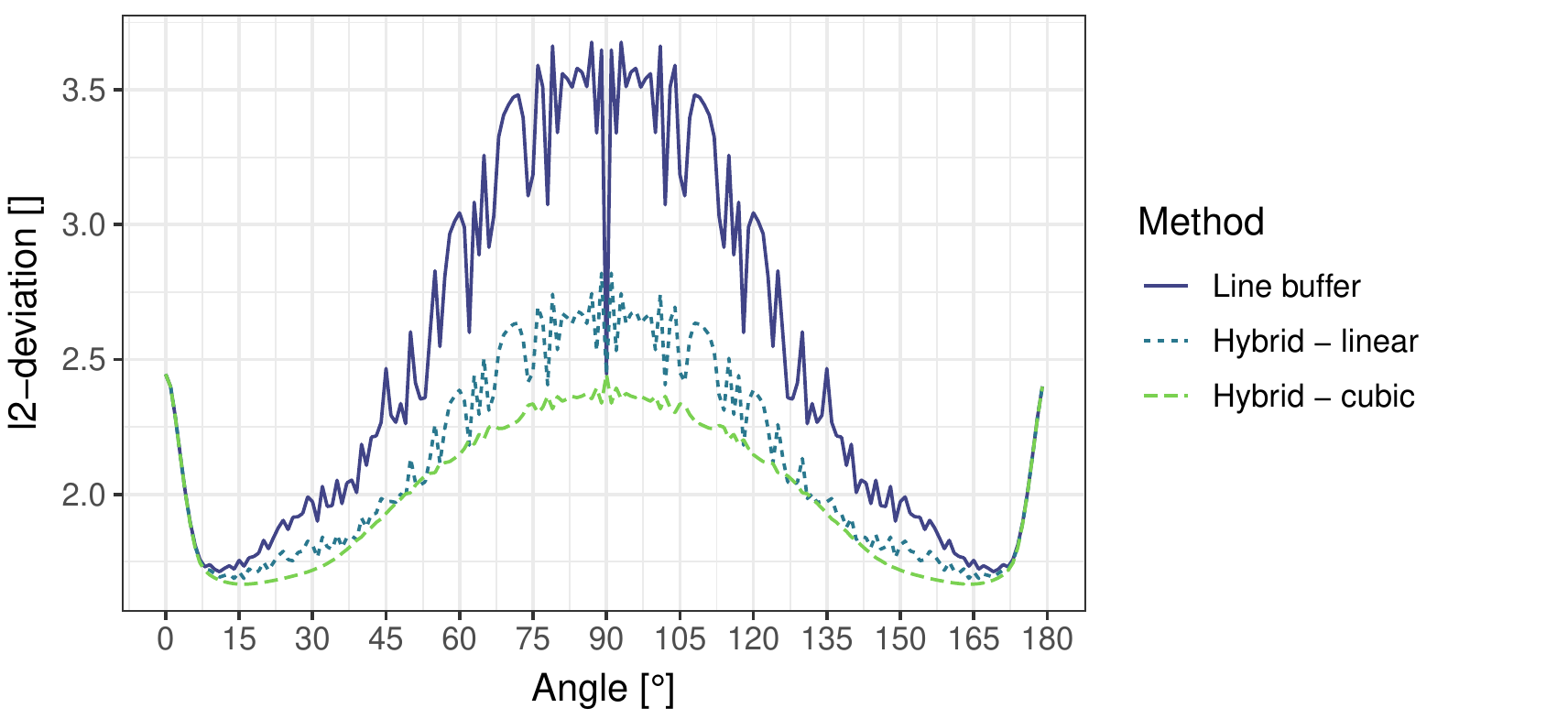}
    \caption{$l^2$-error {in $e^{-3}$} between the reconstructed and the true Gaussian kernel with {$\sigma_1 = 25, \sigma_2 = 2$}, over all angles for the line buffer and the hybrid algorithm with linear resp.~cubic interpolation.}\label{fig:kernelError}
\end{subfigure}

\begin{subfigure}{\textwidth}
    \includegraphics[width=\textwidth]{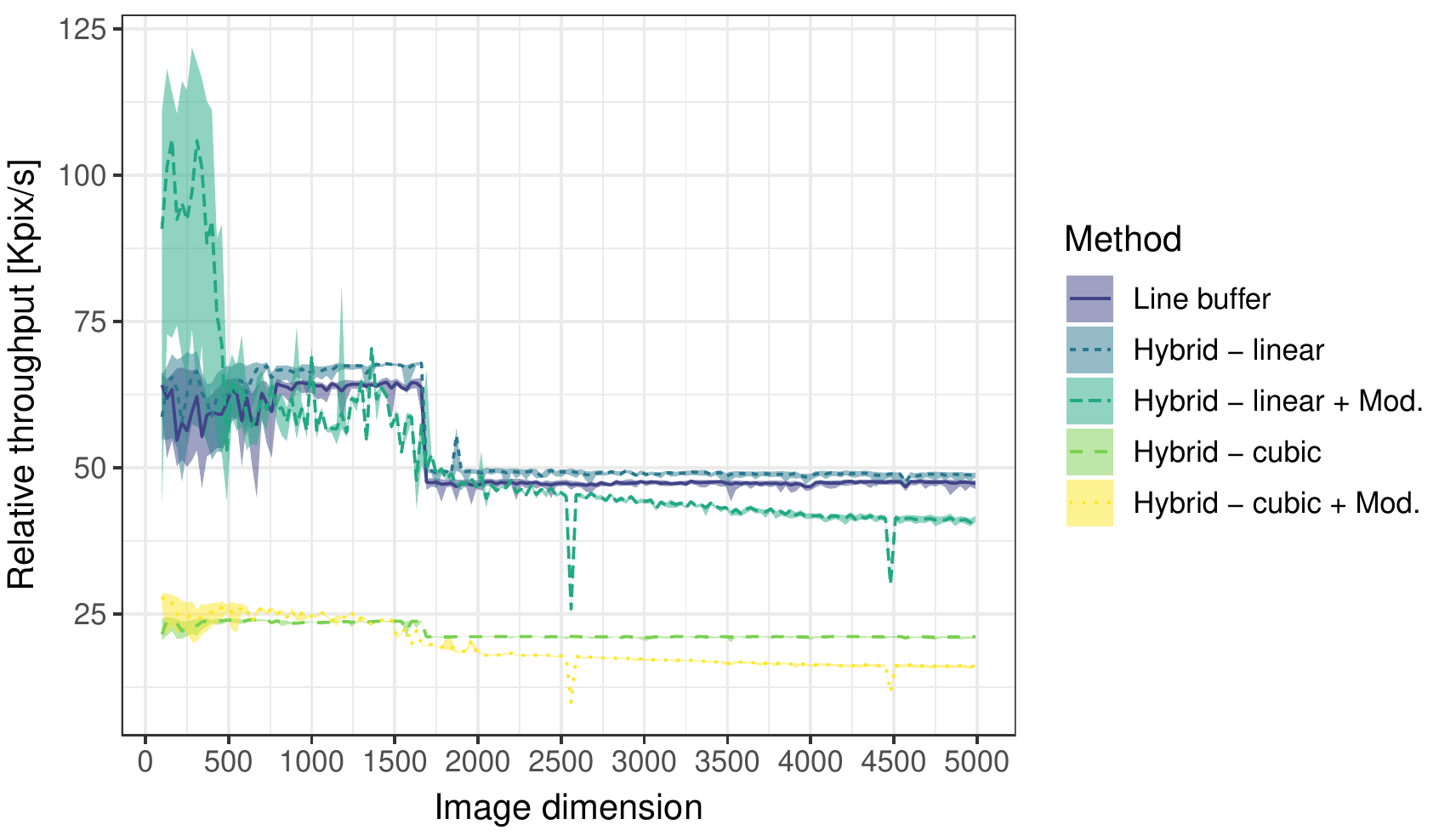}
    \caption{Throughput of the hybrid and line buffer algorithms.}\label{fig:throughput}
\end{subfigure}
\end{figure*}

\subsubsection{Accuracy}
\label{sec:results:aniGauss:accuracy}
We reconstruct the anisotropic Gaussian filter kernel by calculating the unit impulse response, i.e.,~applying the anisotropic Gaussian filter to an image of size $N\,\times\,N$ with $N = 512$, in which all pixel values are $0$ except one pixel in the image center with pixel value $1$.
For each algorithm and variance combination considered here, we compute the $l^2$-deviation between the reconstructed kernel $\pmb{\hat{g}}_\theta$ and the actual kernel $\pmb{g}_\theta$ as a measure of accuracy, i.e.,
\begin{equation}
    \|\pmb{\hat{g}}_\theta - \pmb{g}_\theta \|_{l^2} = \left(\sum_{i, j=1}^N (\hat{g_\theta }_{ij} - {g_\theta }_{ij})^2 \right)^{1/2}.
\end{equation}
The mean and maximum deviation for the rotation angles $\theta  = 0^\circ, 1^\circ, ..., 179^\circ$ are reported in  Table~\ref{tab:errorGaussianKernel}.

The hybrid algorithm with linear interpolation yields more accurate results than the line buffer algorithm.
Cubic interpolation is even more accurate, except for $\sigma_1 = 7.0, \sigma_2 = 4.0$. This is most likely due to ringing artifacts, i.e., oscillations of the interpolation kernel, which is a known problem of cubic interpolation \cite{lehmann99} also known as the Runge phenomenon \cite{gautschi11}, or, more generally, Gibb's phenomenon \cite{hou78}. However, cubic interpolation improves the approximations substantially for variance combinations that otherwise yield comparably large errors for linear interpolation. Note that smaller variances go along with larger errors because the Gaussian approximation is less precise there, see \cite{young95}.

The major-axis modification achieves even higher precision compared to its counterpart without modification. Notably, the hybrid algorithm with linear interpolation and major-axis modification often outperforms the hybrid non-modified algorithm with cubic interpolation.

\begin{table*}[!t]
\caption{$l^2$-deviation in $10^{-3}$ between the reconstructed and the true Gaussian kernel. Mean over all angles $\theta$, maximal error in brackets.\label{tab:errorGaussianKernel}}
\begin{center}
\begin{tabular}{|l|l|
r @{\extracolsep{0pt}}l@{\extracolsep{4pt}}r@{\extracolsep{0pt}}l|
r @{\extracolsep{0pt}}l@{\extracolsep{4pt}}r@{\extracolsep{0pt}}l|
r @{\extracolsep{0pt}}l@{\extracolsep{4pt}}r@{\extracolsep{0pt}}l|
r @{\extracolsep{0pt}}l@{\extracolsep{4pt}}r@{\extracolsep{0pt}}l|
r @{\extracolsep{0pt}}l@{\extracolsep{4pt}}r@{\extracolsep{0pt}}l|}
\hline
  $\sigma_1$ & $\sigma_2$ &\multicolumn{4}{c|}{Line buffer} & \multicolumn{8}{c|}{Hybrid} & \multicolumn{8}{c|}{Hybrid + Mod.}\\
    &&\multicolumn{4}{c|}{Linear}& \multicolumn{4}{c|}{Linear} & \multicolumn{4}{c|}{Cubic} &\multicolumn{4}{c|}{Linear} & \multicolumn{4}{c|}{Cubic}\\
  \hline
  \phantom{1}$2.0$ & $1.0$ & 38&.9 &(60&.8)&  29&.7 &(39&.9)& \textbf{23}&\textbf{.6} &\textbf{(28}&\textbf{.2)} &28&.0 &(30&.0)&25&.4 &\textbf{(28}&\textbf{.2)}\\  
  \hline
  \phantom{1}$5.0$ & $2.0$ &  7&.2 &(10&.6)  & 6&.2 &(7&.8)& 5&.8 &\textbf{(6}&\textbf{.3)}&5&.9 &(6&.5)&\textbf{5}&\textbf{.7} &\textbf{(6}&\textbf{.3)}\\
  \hline
  \phantom{1}$7.0$ & $2.0$ &  5&.7 &(8&.0)& 4&.9 &(6&.0)& 4&.6 &\textbf{(5}&\textbf{.1)}&4&.6 &(5&.2)&\textbf{4}&\textbf{.5} &\textbf{(5}&\textbf{.1)}\\  
  \phantom{1}$7.0$ & $4.0$ &  2&.8 &(2&.9)& 2&.7 &(2&.8)&\textbf{2}&\textbf{.6} &(3&.1)&2&.7 &\textbf{(2}&\textbf{.7)}&\textbf{2}&\textbf{.6} &(3&.2)\\  
  \hline
  $10.0$ & $0.5$ & 35&.7 &(75&.7) & 23&.1 &(60&.8) & 14&.3 &(30&.4) &16&.9& (29&.0)&\textbf{12}&\textbf{.0} &\textbf{(18}&\textbf{.3)}\\  
  $10.0$ & $1.25$ & 9&.5 &(17&.5)& 7&.2 &(11&.4)& 5&.9 &\textbf{(8}&\textbf{.3)}&5&.6 &\textbf{(8}&\textbf{.3)}&\textbf{5}&\textbf{.4} &\textbf{(8}&\textbf{.3)} \\ 
  $10.0$ & $2.0$ & 4&.5 &(7&.0)& 3&.9 &(4&.9)& 3&.6 &\textbf{(4}&\textbf{.1)}&3&.6 &\textbf{(4}&\textbf{.1)}&\textbf{3}&\textbf{.5} &\textbf{(4}&\textbf{.1)} \\
  \hline
  $20.0$ & $0.5$ &  24&.6 &(44&.0)& 15&.8 &(37&.3)& 9&.8& (19&.4)&10&.9 &(22&.6)&\textbf{7}&\textbf{.7} &\textbf{(12}&\textbf{.8)}\\ 
  $20.0$ & $1.25$ & 6&.1 &(10&.4)& 4&.6 &(7&.6)& 3&.9 &\textbf{(5}&\textbf{.8)}&3&.4 &\textbf{(5}&\textbf{.8)}&\textbf{3}&\textbf{.3} &\textbf{(5}&\textbf{.8)}\\
  $20.0$ & $2.0$ & 2&.9 &(4&.2)&2&.4 &(3&.2)& 2&.3 &\textbf{(2}&\textbf{.8)}&2&.2 &\textbf{(2}&\textbf{.8)}&\textbf{2}&\textbf{.1} &\textbf{(2}&\textbf{.8)}\\ 
  \hline
  $25.0$ & $0.5$ &  21&.8 &(37&.9)& 13&.9 &(31&.4)& 8&.7 &(16&.7) &9&.6 &(15&.7)&\textbf{6}&\textbf{.6} &\textbf{(11}&\textbf{.5)}\\ 
  $25.0$ & $1.25$ & 5&.5 &(9&.3)&4&.1 &(6&.6)& 3&.4 &(5&.2)&2&.9 &(5&.2)&\textbf{2}&\textbf{.8} &\textbf{(5}&\textbf{.1)}\\ 
  $25.0$ & $2.0$ &  2&.5 &(3&.7) &2&.1 &(2&.8)& 2&.0& \textbf{(2}&\textbf{.4)} &\textbf{1}&\textbf{.8} &\textbf{(2}&\textbf{.4)}&\textbf{1}&\textbf{.8} &\textbf{(2}&\textbf{.4)}\\ 
   \hline
\end{tabular}
\end{center}
\end{table*}

For elongated Gaussian kernels, the $l^2$-error changes considerably over all rotations: It is lowest for small angular deviations from the $x$-axis, that is, $\theta=0^\circ$. Between $50^\circ$ to $130^\circ$ it is considerably larger, peaking around $90^\circ$. This conforms with our motivation for the major-axis modification{. }
Employing the hybrid algorithm, the deviations shrink significantly, see Fig.~\ref{fig:kernelError}.

\subsubsection{Throughput}
We test the algorithms' data throughput by applying the filter $50$ times to Gaussian noise images of sizes $N \times N$ with $N = 100, 130, ..., 4\,990$ and calculate the trimmed mean excluding top and bottom $10\%$.
Fig.~\ref{fig:throughput} shows that the hybrid algorithm with linear interpolation is slightly faster than the line buffer algorithm, at least for larger image sizes.
Cubic interpolation, however, takes considerably more time. This conforms to the theoretical results {above.} 

The major-axis modification further slows down the algorithms. For $45^\circ \leq \theta \leq 135^\circ$, the filter in $\nu_\ast$-direction iterates over all image columns, while the image pixels are saved adjacently within a line. Therefore, memory access is more expensive than it is without modification, the more so, the larger the image. This can be circumvented at the cost of memory by saving the image adjacently within a column for $45^\circ \leq \theta \leq 135^\circ$.

For all three algorithms without modification, there are two different speed plateaus in the throughput, see Fig.~\ref{fig:throughput}. As Lampert \& Wirjadi~\cite{lampert06} argue, the throughput is dependent on the image size: In our implementations --- for the hybrid as well as the line buffer algorithm --- we use 4 buffers for the filter history while reading from and writing into the same image. For small image dimensions, these buffers fit into the CPU's L1 data cache. For larger image sizes, the buffer sizes exceed the cache size slowing down the computations. In our test setup the drop at $N = 1\,690$ corresponds with the system's size of the L1 data cache, namely 64 KiB, see Fig.~\ref{fig:throughput}.

\subsection{Experimental Validation of the MR Method}
\label{subsec:ExperimentalValidation:maxResponse}
\begin{figure*}
    \centering
    \begin{tikzpicture}[scale=1.25]
        \draw [thick](3.0,0) -- (-5,0) node [left] { $c$};
        \draw [thick](3.02,0.10) -- (2.98,-0.1);
        \draw [thick](3.12,0.10) -- (3.08,-0.1);
        \draw [thick](5,0) -- (3.1,0);
        
        \draw (-5,1.3) node[inner sep=0pt] {\includegraphics[width=.15\linewidth]{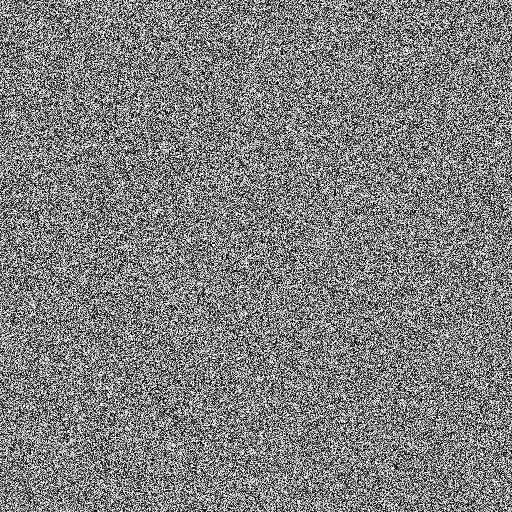}};
        \draw [thick](-5,0.10) -- (-5,-0.1) node [below] {$0$};

        \draw (-1.25,1.3) node[inner sep=0pt] {\includegraphics[width=.15\linewidth]{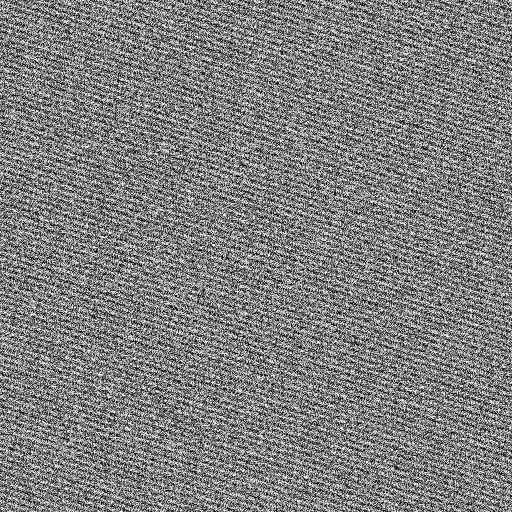}};
        \draw [thick](-1.25,0.10) -- (-1.25,-0.1) node [below] { $0.25$};
        
        \draw (1.0,1.3) node[inner sep=0pt] {\includegraphics[width=.15\linewidth]{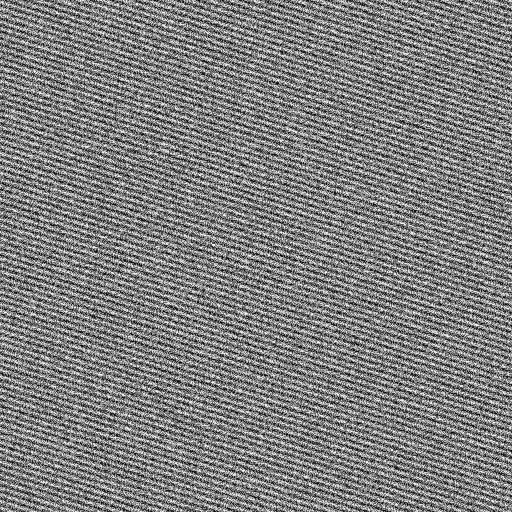}};
        \draw [thick](1.0,0.10) -- (1.0,-0.1) node [below] { $0.4$};
        
        \draw (5,1.3) node[inner sep=0pt] {\includegraphics[width=.15\linewidth]{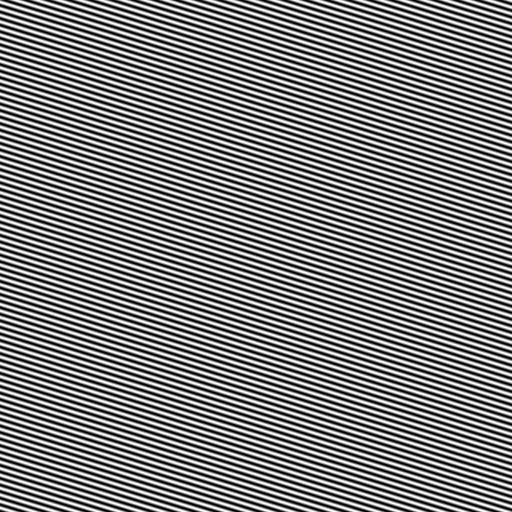}};
        \draw [thick](5,0.10) -- (5,-0.1) node [below] { $1$};
        
    \end{tikzpicture}
    \caption{Visualization of the experimental data set for varying contrast $c$.}\label{fig:contrastViz}
\end{figure*}

\begin{figure*}%
    \centering
    \includegraphics[width=\textwidth]{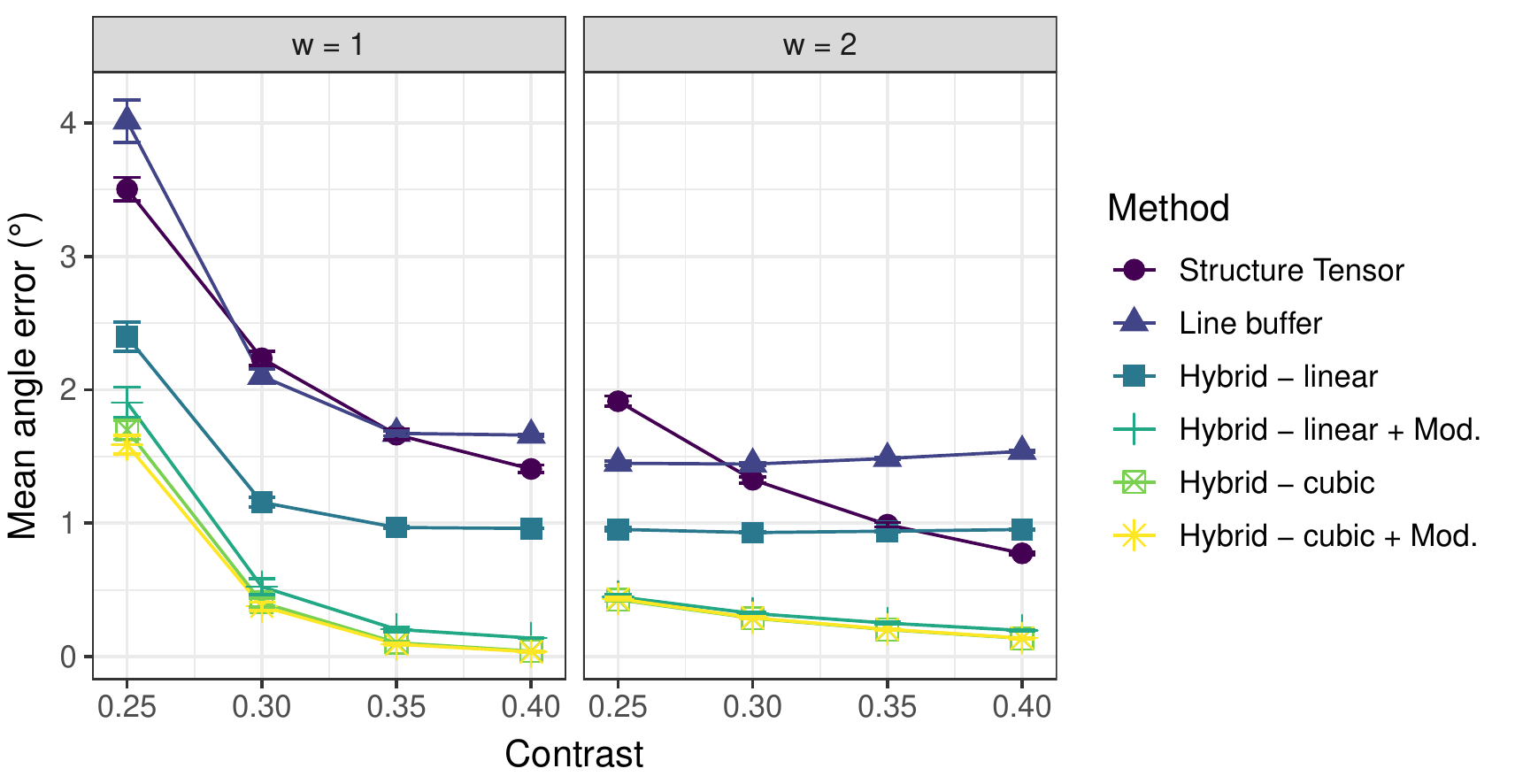}
    \caption{Mean angle error for 50 noise images overlayed with synthetic fiber images with direction $\theta = 0^\circ, ..., 179^\circ$, and with varying contrast. For each noise image contrast combination, the MAE's maximum over fiber directions was calculated. The mean and standard deviations over 50 noise images are depicted as point symbol with bars for each contrast and algorithm. Note, however, that the standard deviations are small and therefore the bars delimiting the interval are in many cases covered by the symbol for the mean value.}
    \label{fig:contrastError}
\end{figure*}
The experiments in the previous subsection have shown that anisotropic Gaussian filters are generally more accurately calculated with the hybrid than with the line buffer algorithm, especially with the major-axis modification. This section will show that these results translate to the accuracy of the MR method.

\subsubsection{Setup}
\label{subsubsec:SetupMaxResponse}
We evaluate the MR method on synthetic images with known constant fiber orientation. The design of the images is inspired by our application example. There, bundles of nearly parallel thin fibers form the main building block of the microstructure. The synthetic images shall mimic the fiber system within one such bundle.

Given an image of size $512\,\times\,512$ pixels, a width parameter $w$, and an angle $\theta$, we define an image $F_{\theta, w}$ by setting
\begin{equation}
    F_{\theta, w}(x, y) = \frac{\sin(x \sin(\theta) + y \cos(\theta))}{2w} + \frac{1}{2}.
\end{equation}
For each $\theta = 0^{\circ}, 1^{\circ}, 2^{\circ}, ..., 179^{\circ}$ and $w = 1, 2$, we generate such an image. These gray-value images represent idealized fiber bundles with known fiber direction and a radius of $r = \pi w/2$\,pixels. {Note that we choose a significantly smaller radius than Wirjadi et~al.~\cite{wirjadi16} and Pinter et~al.~\cite{pinter18}., who use cylinders with a radius of at least 2.5\,voxels. Our choice is again inspired by our application example.}

As background noise, we generate images $B$ of size $512\,\times\,512$ with pixels sampled from the uniform distribution in $[0, 1]$. 

To model images of varying contrast between background and fiber, we consider images of the form $(1 - c) B + c F_{\theta, w}$ for $c \in [0,1]$, see Fig.~\ref{fig:contrastViz}. 
The MR method is applied as described below, which yields a mean absolute angular error \emph{MAE} w.r.t.~the known fiber direction. For comparison, we additionally apply the algorithm to images preprocessed by a median filter of size $3\,\times\,3$. This is motivated by the fact that smoothing with median filters is a typical preprocessing step for real image data.

The MAE is determined as follows. The synthetic fiber image $F_{\theta, w}$ is binarized with a threshold of $0.75$. The resulting image serves as a mask to include only pixels within fiber cores.

Further, we want to make sure that we only estimate the estimating bias and do not confound it with a sampling bias, i.e.,~lower or higher sampling probability for certain directions. For example, in an image, one can place more fibers and longer fibers in diagonal directions than in the horizontal and vertical direction. Therefore, we only evaluate pixels within a circle around the image's center. In order to avoid boundary effects, the circle radius is set to $206$~pixels.

For each realization $B$, we are interested in the maximum of the MAE over all $\theta$ per method. We run the MR method with $\sigma_1 = 20.0, \sigma_2 = 0.75w {\approx {r}/{2}}$
{, the structure tensor with $\sigma = r, \ \rho = 6.0$ and the Hessian matrix with $\sigma = r$ as argued in the Methods section.}

\subsubsection{Results}

\begin{figure*}
    \begin{subfigure}{\textwidth}
        \centering
        \includegraphics[width=\textwidth]{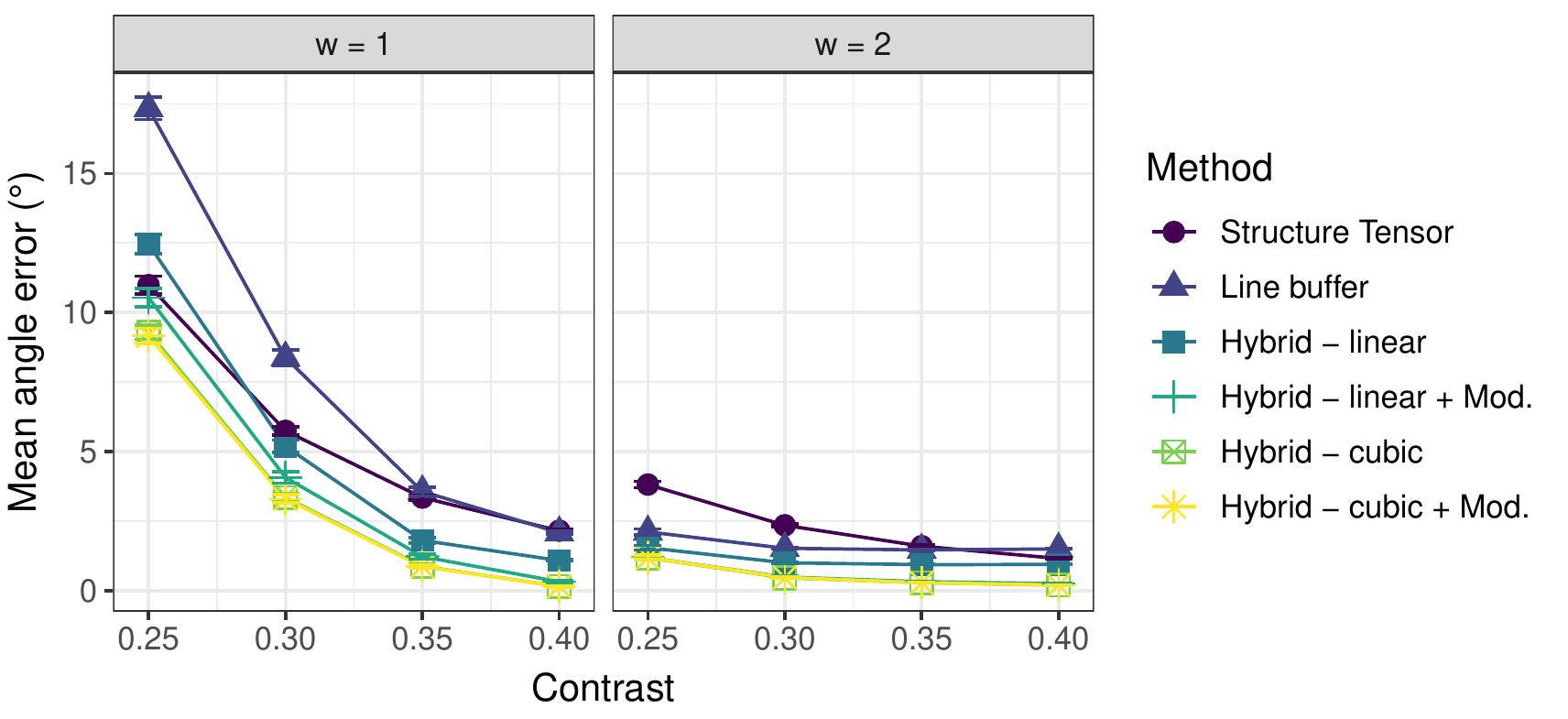}
        \caption{Experimental results after applying the median filter.}
    \end{subfigure}
    \begin{subfigure}{\textwidth}
        \centering
        \includegraphics[width=\textwidth]{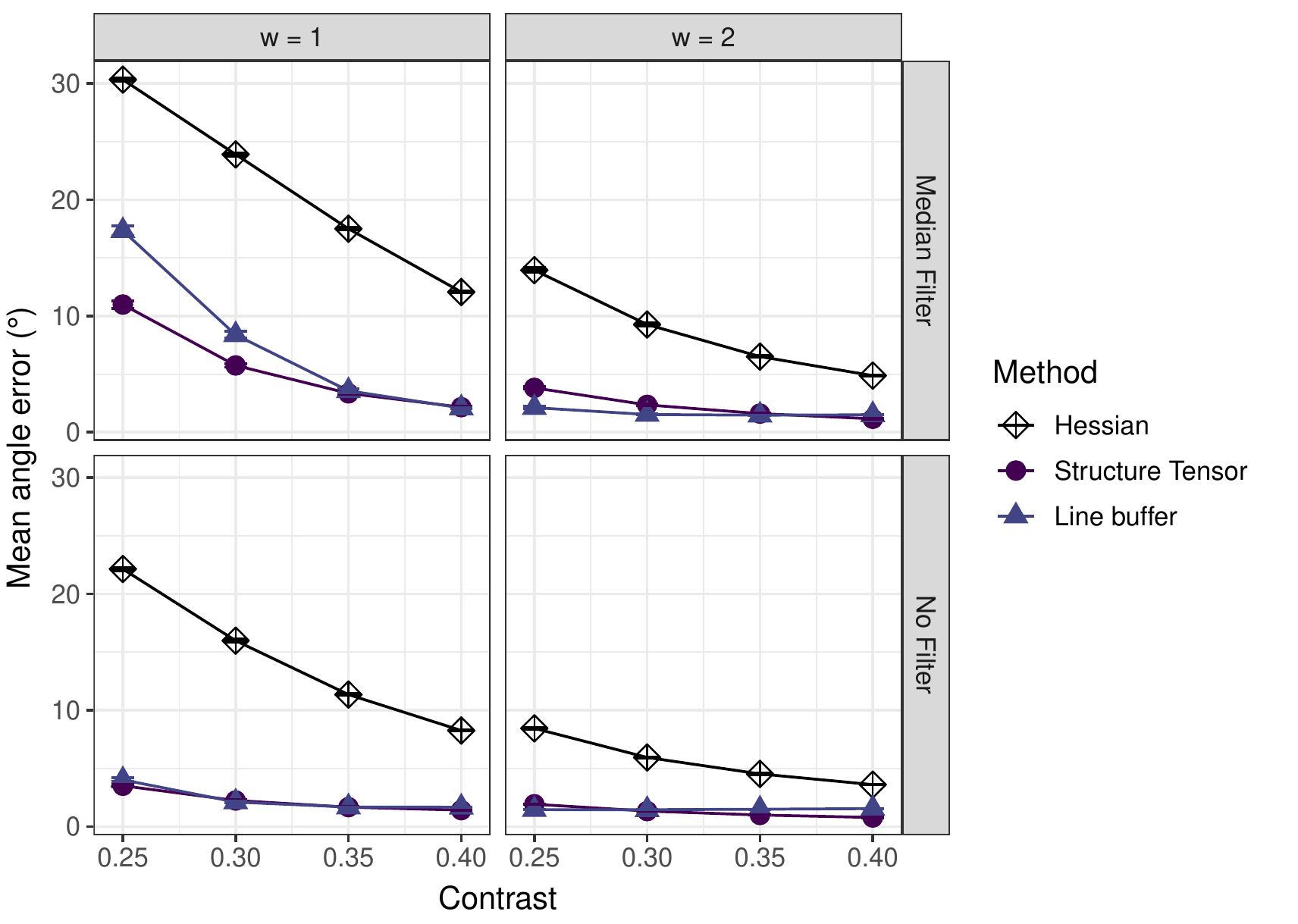}
        \caption{Experimental results for the Hessian matrix.}
    \end{subfigure}
    \caption{Mean angle error for 50 noise images overlayed with synthetic fiber images with direction $\theta = 0^\circ, ..., 179^\circ$, and with varying contrast. For each noise image contrast combination, the MAE's maximum over fiber directions was calculated. The mean and standard deviations over 50 noise images are depicted as point symbol with bars for each contrast and algorithm. Note, however, that the standard deviations are small and therefore the bars delimiting the interval are in most cases covered by the symbol for the mean value.}
    \label{fig:contrastErrorMedian}
\end{figure*}

Fig.~\ref{fig:contrastError} shows the mean error for unfiltered images 
and the standard deviations for 50 noise realizations. Considering the MR method, the hybrid algorithm outperforms the line buffer algorithm, especially for low-contrast cases. The cubic interpolation is more accurate than the linear interpolation, especially for the high contrast, but also for the low-contrast setup. The hybrid algorithm with linear interpolation and major-axis modification performs nearly as well as it does with cubic interpolation. For cubic interpolation, however, it performs just as well with the major-axis modification as it does without it. {The effect of low contrast is reduced for the larger fiber diameter of $r = \pi$ in comparison to $r = {\pi}/{2}$.

The structure tensor, however, is affected by low contrast for both $r = {\pi}/{2}, \pi$. For $r = {\pi}/{2}$, it is about as accurate as the MR method using the line buffer algorithm, so it is outperformed by the hybrid algorithm. For $r = \pi$, it is always outperformed by the MR method with a hybrid algorithm and modification. Note that for higher contrast, it still performs better or as well as the MR method using the line buffer algorithm, but for the lowest contrast, it is outperformed by all algorithms employed in the MR method.

The Hessian matrix is not able to match the performance of the other methods, see Fig.~\ref{fig:contrastErrorMedian}(b). }

Applying a median filter to noisy images is a common preprocessing step to get rid of noise while preserving edges. However, the errors are considerably larger than for unfiltered images as direction information is lost by the undirected median filter, see Fig.~\ref{fig:contrastErrorMedian}(a).

\subsection{Application to Sheet Moulding Compounds}
\label{sec:Application}
In this section, we apply the MR method to low-contrast image data of sheet molding compound materials. Sheet molding compounds (\textit{SMC}) are a type of material consisting of stacked layers of fibers. In the automotive industry, SMC are of high interest due to their versatile behavior such as light weight, high stiffness, and strength, which is determined by their fiber direction distribution \cite{orgeas2011sheet}. Computed tomography imaging of SMC is challenging due to the high fiber volume fraction and the low difference in X-ray absorption of fiber and matrix material. {In the following, we will use the MR method both for fiber direction estimation and fiber enhancement, which is useful for segmenting the fibers.}

\subsubsection{Sheet Molding Compound with Glass Fibers}
\label{sec:ALMA}

\begin{figure}[t]%
\begin{subfigure}{0.45\textwidth}
    \includegraphics[width=\linewidth]{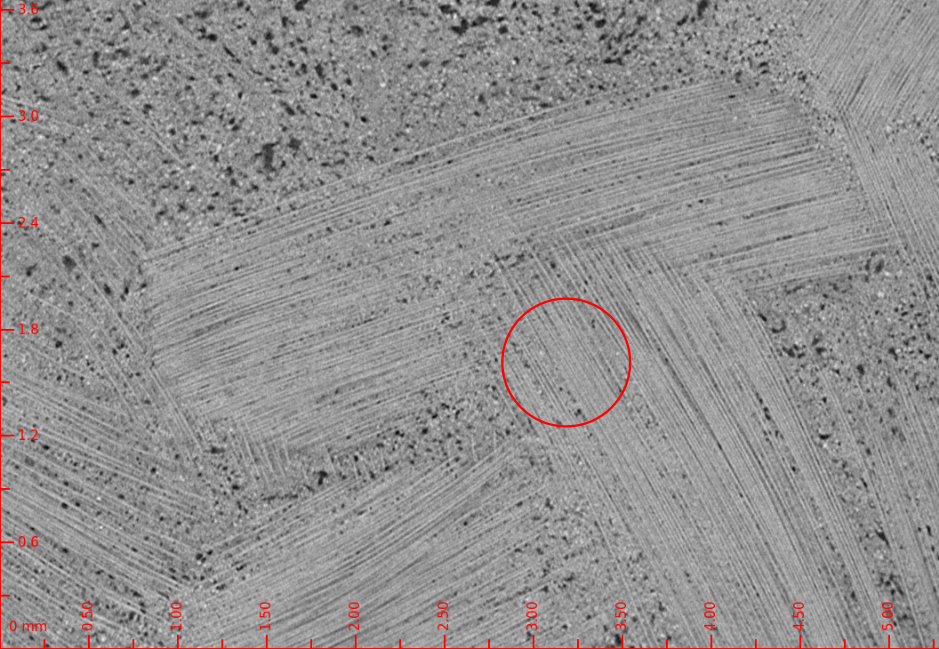}
    \caption{Original slice, gray-values are spread for improved visibility.}
\end{subfigure}
\begin{subfigure}{0.45\textwidth}
    \includegraphics[width=\linewidth]{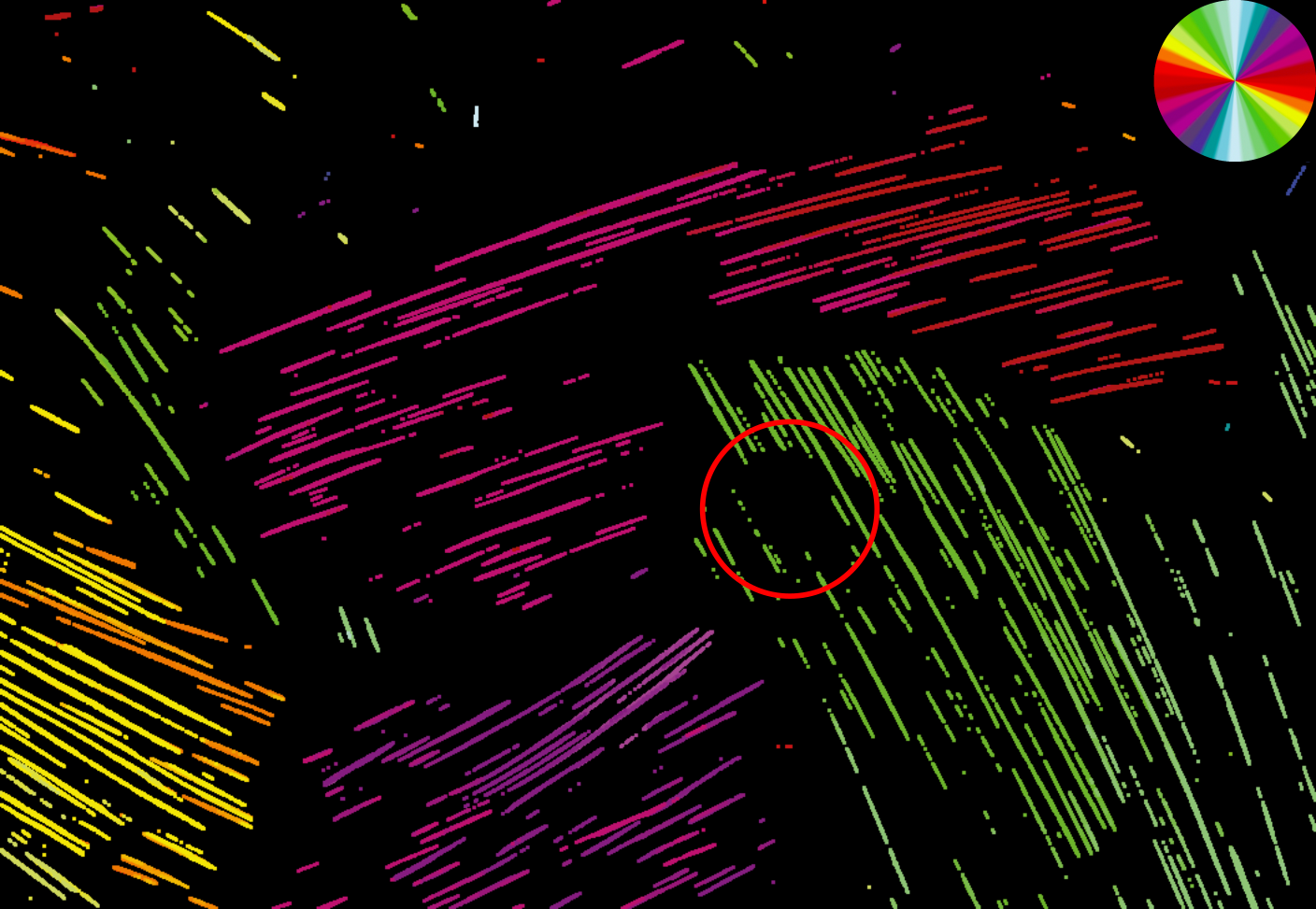}
    \caption{Resulting slice with the line buffer algorithm.\\~}
\end{subfigure}

\begin{subfigure}{0.45\textwidth}
    \includegraphics[width=\linewidth]{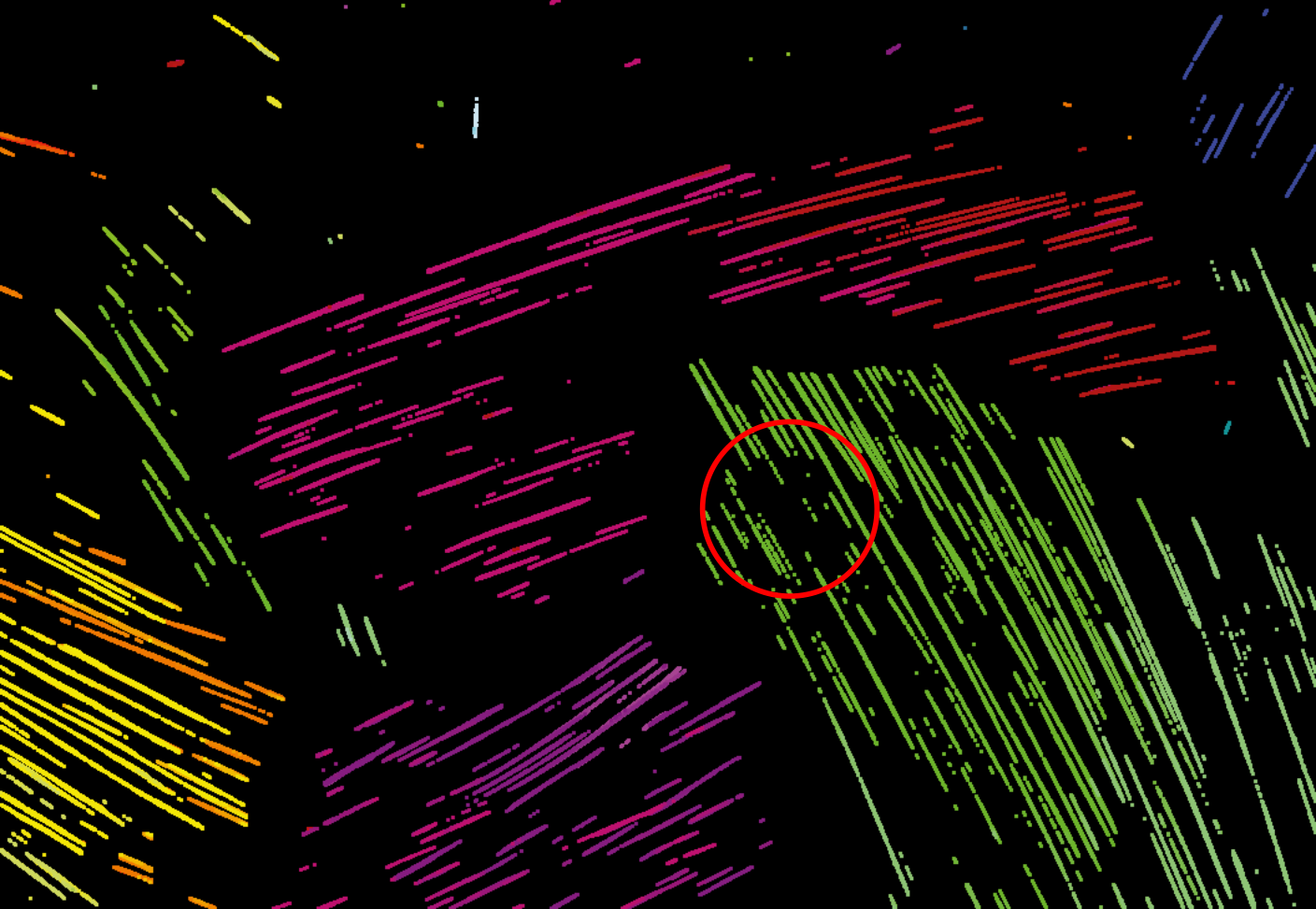}
    \caption{Resulting slice with the hybrid algorithm using linear interpolation and the modification.}
\end{subfigure}
\begin{subfigure}{0.45\textwidth}
    \includegraphics[width=\linewidth]{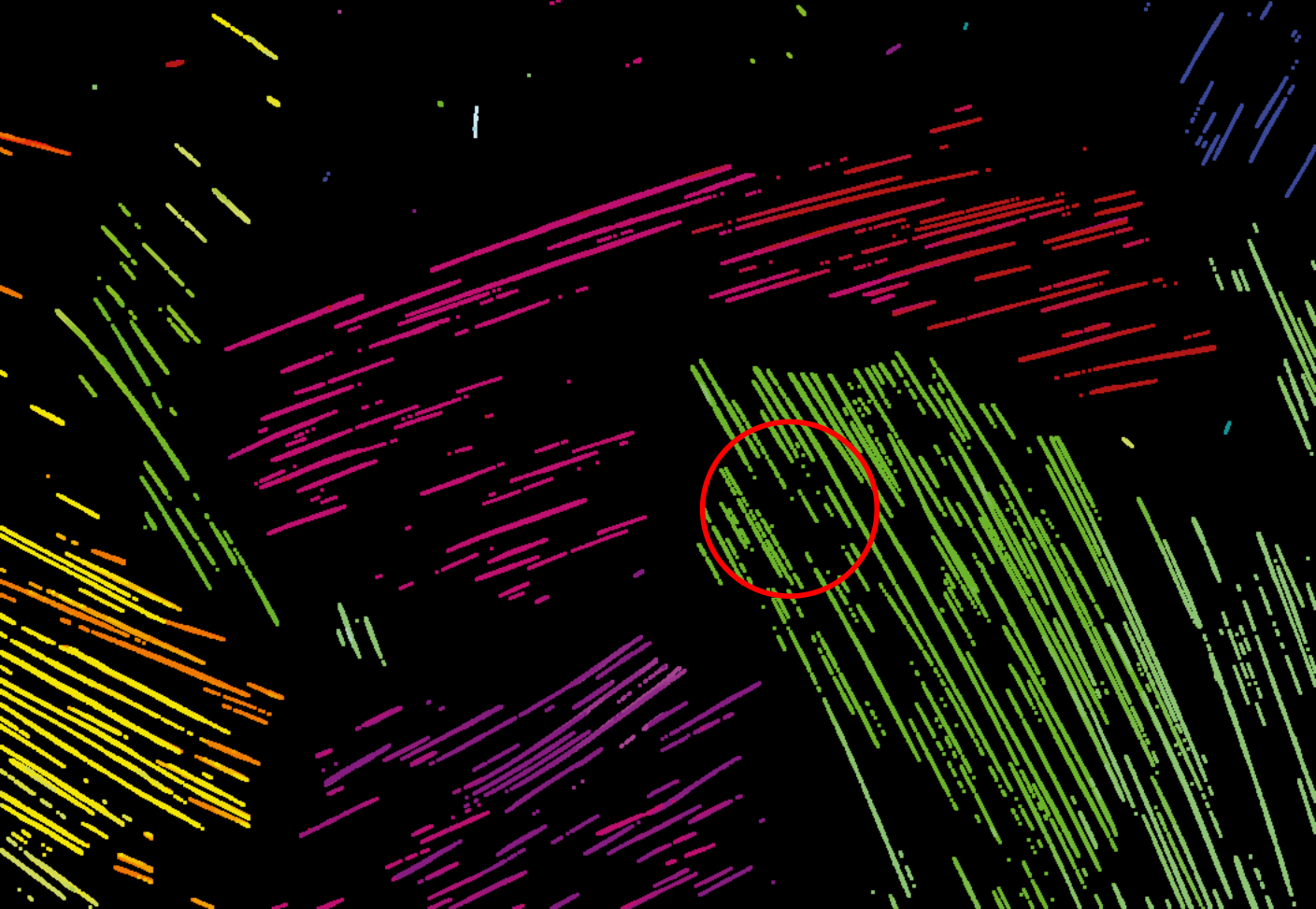}
    \caption{Resulting slice with the hybrid algorithm using cubic interpolation.}
\end{subfigure}
\caption{ \label{fig:application:imageALMA} Analysis of SMC with glass fibers using the MR method with $\sigma_1 = 20.0$, $\sigma_2 = 0.5$, and binarization.}
\end{figure}

First, we consider an SMC material with glass fibers, see Fig.~\ref{fig:application:imageALMA}.
The image was taken using the \textmu CT device at the Fraunhofer ITWM, Kaiserslautern, Germany, with a voltage of $120$\,kV, an integration time of $999$\,ms, and 1\,200\,projections/angular steps. The device uses a Feinfocus FXE-225 X-ray tube and a PerkinElmer detector with $2\,048\,\times\,2\,048$\,pixels \cite{uCT2022}. As specified by the manufacturer, the fibers' diameter is $10$\,\textmu m. The material was scanned with a pixel spacing of $5$\,\textmu m, deliberately undersampling the fibers for the sake of imaging representative sample volumes.

{Following the arguments from the Methods section,} we applied the line buffer and the hybrid algorithm with both linear and cubic interpolation to the sample {with $\sigma_1 = 0.5, \sigma_2 = 20.0$. For the structure tensor and the Hessian matrix, we used $\sigma = 1.0$ and $\rho = 6.0$. Additional to the direction estimation, the MR method supplies a fiber enhancement with its maximal response. Combined with Frangi~\textit{et~al.}'s~\cite{Frangi98multiscalevessel} enhancement filtering, we employ it to segment the fiber system. Subsequently, we add a postprocessing step based on Sliseris~\textit{et~al.}'s~\cite{sliseris15} work.
For further details see \cite{alma-paper}.

For these images, we do not have a ground truth, but visually, the masks produced by the MR method all appear quite accurate, see Fig.~\ref{fig:application:imageALMA}. 
Note that for very diffuse fiber bundles, the hybrid algorithm using cubic interpolation looks the most accurate. This result is consistent with its higher accuracy for the response to a unit impulse
: The higher accuracy very likely translates to the response to fibers, which makes the enhanced fibers even more distinguishable, the basis of our segmentation approach.}


\subsubsection{Sheet Molding Compound with Carbon Fibers}
\label{sec:IVW}
As a second application, we consider images of the material SMCarbon\textregistered~24 CF50-3K by POLYNT Composites Germany GmbH. It consists of carbon fibers with a length of $25$\,mm within a vinyl ester resin. The fiber diameter is not known directly as it also changes under pressure.

The sample was scanned with the X-ray microscope Xradia~520 by Carl Zeiss Microscopy GmbH \cite{XRadia520} with a pixel spacing of $24.93$\,\textmu m, a voltage of $60$\,keV, a power of $5$\,W, and 3\,201 projections. The exposure time was $2$\,s, where 20 single images were taken with an exposure time of $0.1$\,s and then averaged. For the field-of-view, they used $76$\,mm\,$\times\,48$\,mm.

{We applied our versions of the MR method with $\sigma_1 = 25.0, \sigma_2 = 2.0$.} 
Following Schladitz et~al.~\cite{schladitz16}, we binarized the maximal filter response using Niblack's~\cite{niblack86} local thresholding with a window size of $w = 4 \sigma_2$, and the threshold $0.6$. Further, we excluded components that have a pixel size lower than $100$ after eroding the mask with a square of size $2\,\times\,2$. {For the sake of consistency, we applied the structure tensor and the Hessian matrix with $\sigma = 4.0$ and $\rho = 6.0$.}

Despite there barely being any contrast within the fiber bundles, the MR algorithm provides a fairly accurate estimation of fiber directions{, whereas the gradient-based methods are struggling, see Fig.~\ref{fig:application:imageIVW}. We compared the histograms of different algorithms for the MR method, see supplementary material. Strikingly, the line buffer and the hybrid algorithm using linear interpolation without modification apparently overestimate the direction $90^\circ$ while underestimating the neighboring $89^\circ$ and $91^\circ$. This conforms to the error behavior of the kernel as plotted in Fig.~\ref{fig:kernelError}. The other algorithms show minor deviations from each other.}

\begin{figure*}%
\begin{subfigure}{0.45\textwidth}
    \includegraphics[width=\textwidth]{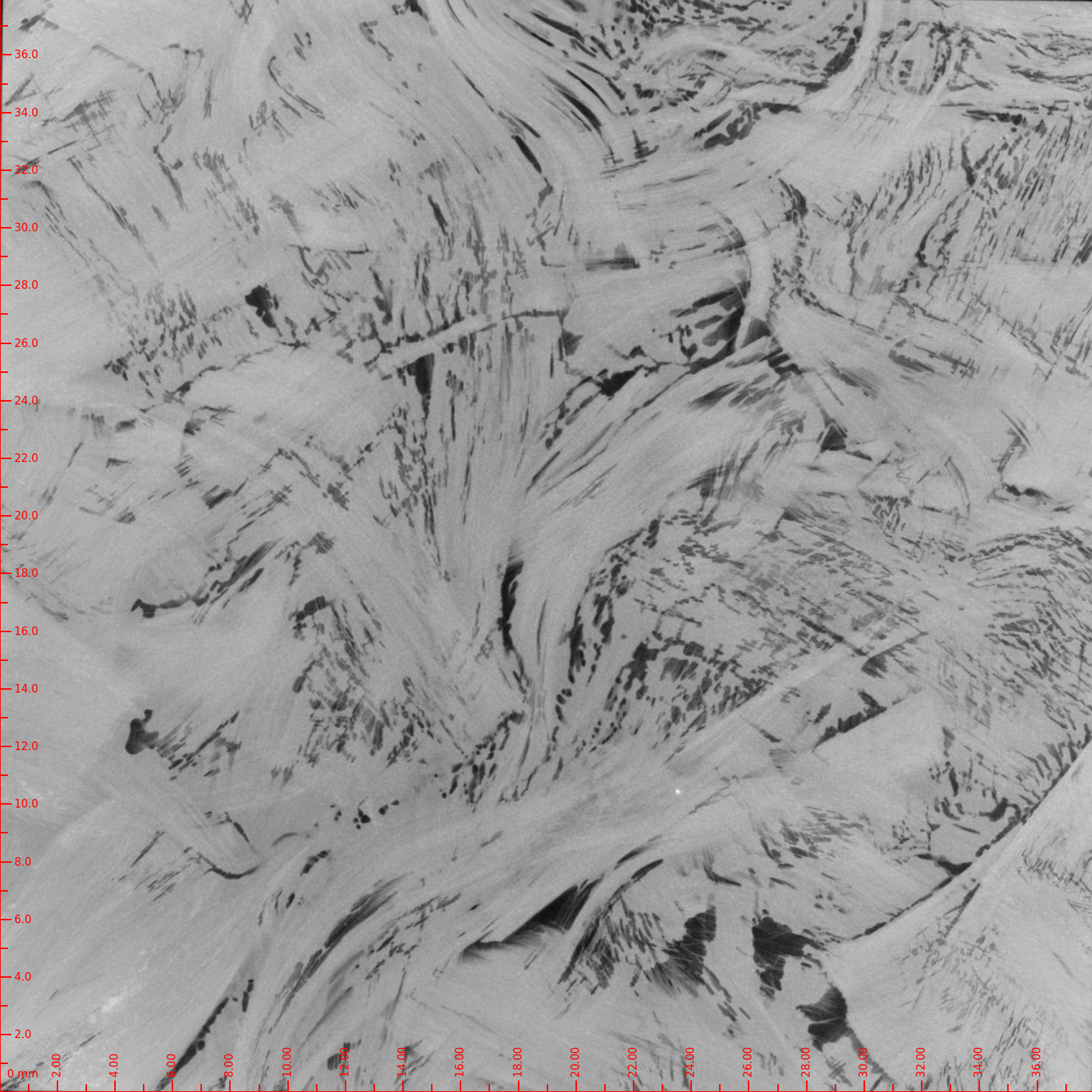}
    \caption{Original slice, gray-values are spread for improved visibility.}
\end{subfigure}
\begin{subfigure}{0.45\textwidth}
    \includegraphics[width=\textwidth]{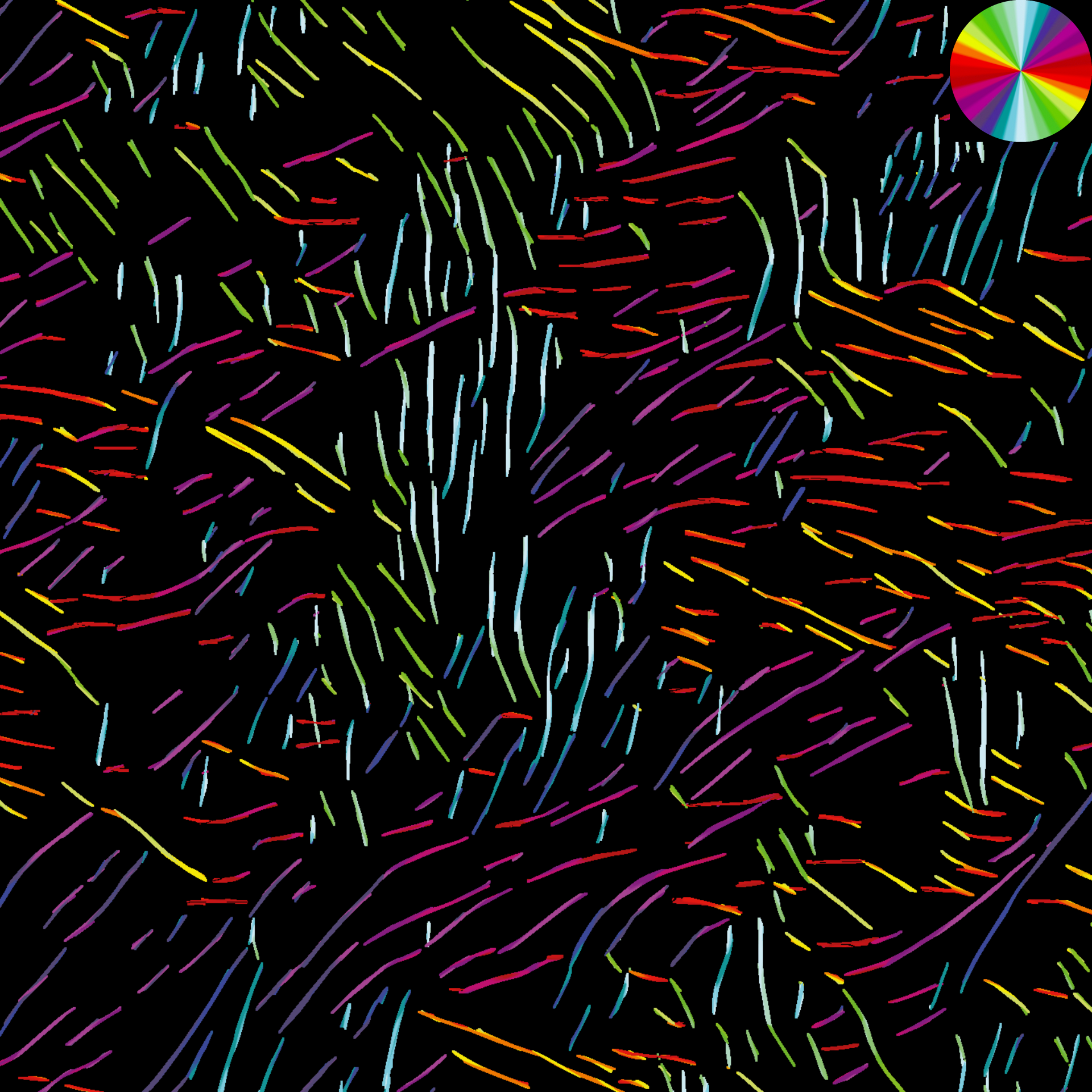}
    \caption{Result using the MR method.\\~}
\end{subfigure}

\begin{subfigure}{0.45\textwidth}
    \includegraphics[width=\textwidth]{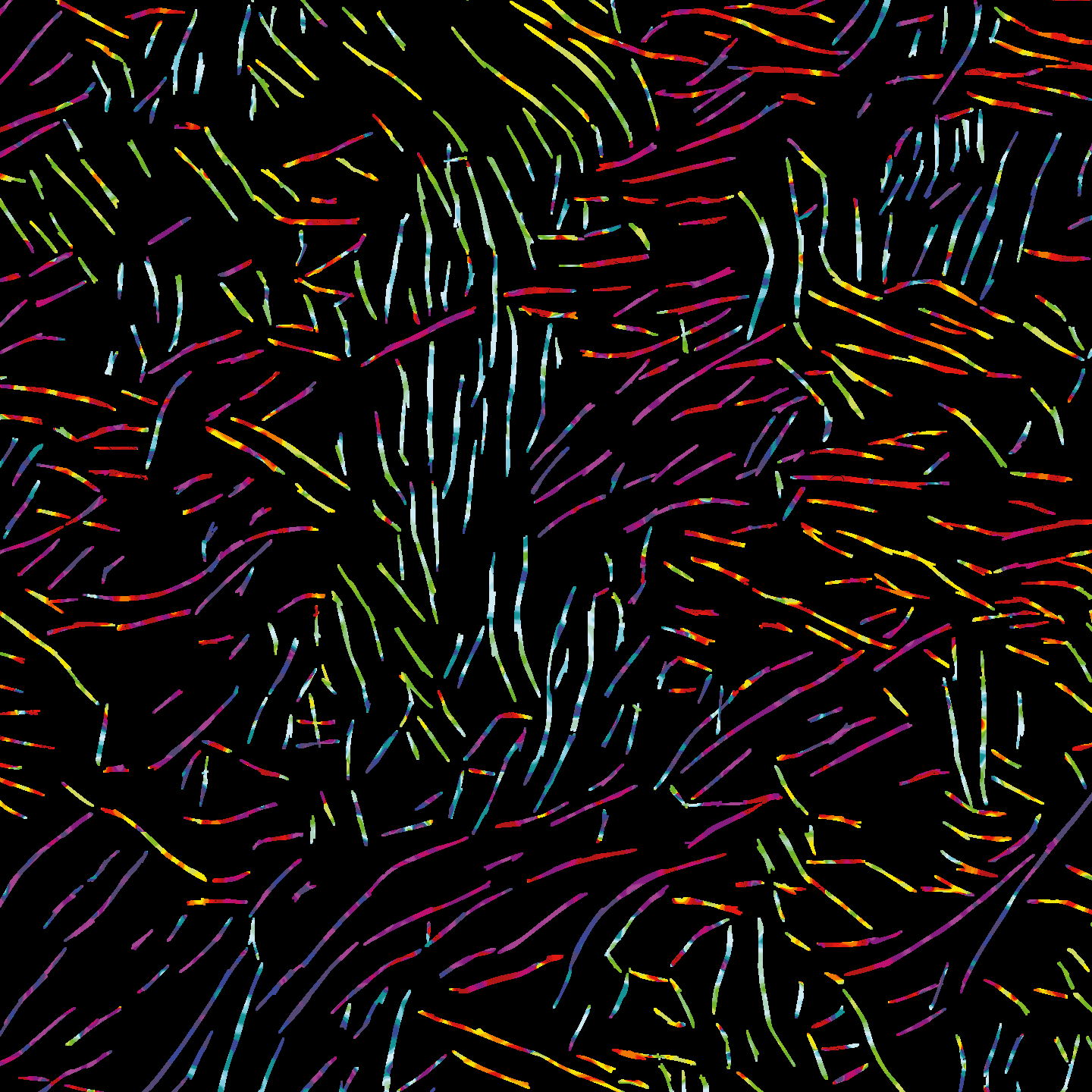}
    \caption{Result using the structure tensor.}
\end{subfigure}
\begin{subfigure}{0.45\textwidth}
    \includegraphics[width=\textwidth]{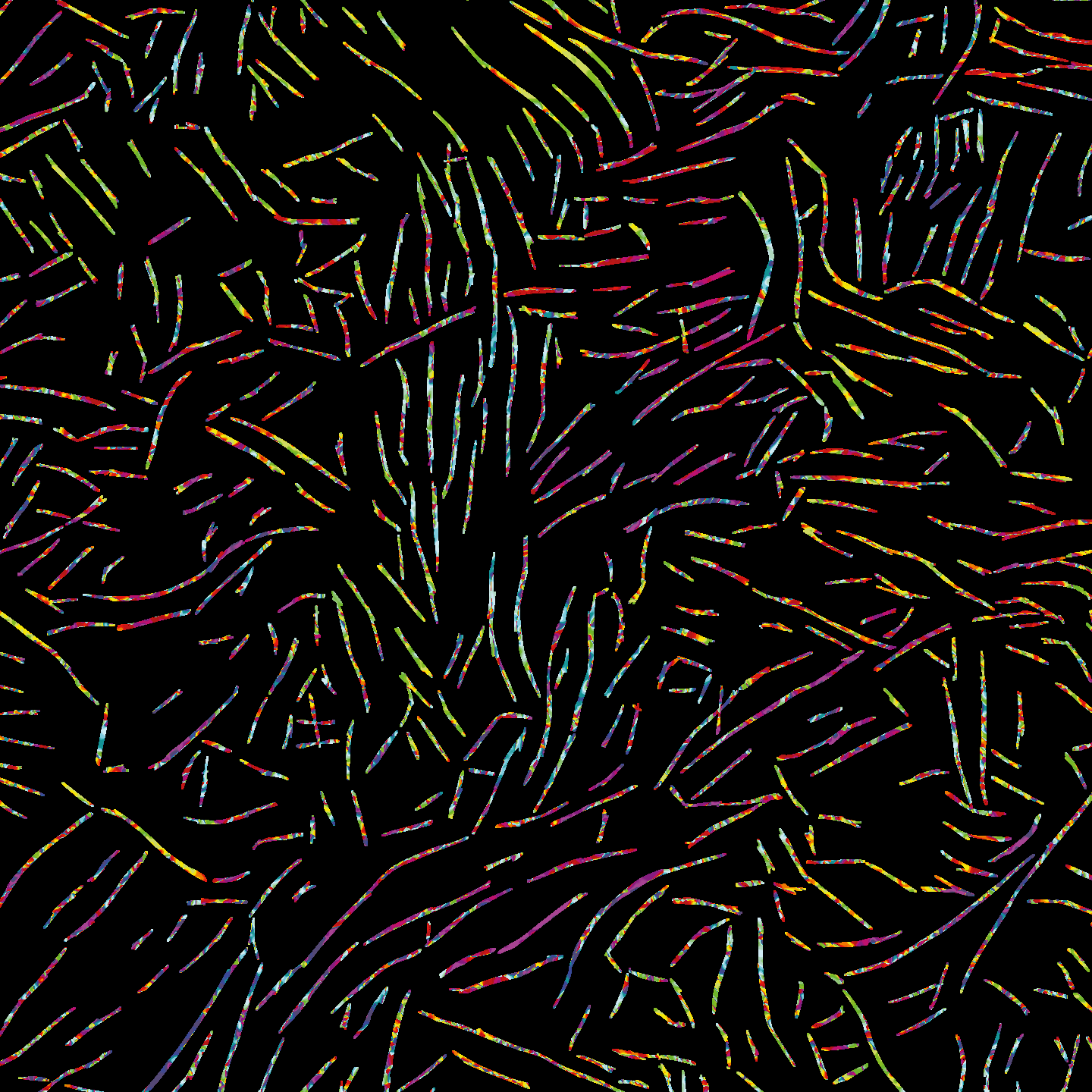}
    \caption{Result using the Hessian matrix.}
\end{subfigure}
\caption{ \label{fig:application:imageIVW} Direction estimation of SMC with carbon fibers. The directions are color-coded, see color wheel at the top-right side.}
\end{figure*}

\section{Discussion}
We proposed an alternative algorithm for elongated anisotropic Gaussian filters in 2D, which improves throughput and accuracy. Employed in a numerical scheme for estimating fiber directions, namely the maximal response of anisotropic Gaussian filters, it improves accuracy, especially for noisy images with low contrast.
{Moreover, it outperforms established methods using the structure tensor or the Hessian matrix on synthetic images of fibers.}

{We present two real-world data sets of sheet molding compounds, to which we apply the method successfully. This application also inspired our experimental setup, for which we generated synthetic images containing straight and parallel fiber bundles. Here, our modifications to the algorithm yield improved precision. Note that the method still performs well on visibly bent carbon fibers.}



\section*{Acknowledgements}
We thank Franz Schreiber, Fraunhofer Institute for Industrial Mathematics (ITWM), for the computed tomography imaging.
This work was supported by the German Federal Ministry of Education and Research under Grant Agreement No:~05M22UKA.

The project {on SMC with glass fibers}
, named ALMA, has received funding from the European Union’s Horizon 2020 Research and Innovation Programme under Grant Agreement No:~101006675.

The computed tomography scans {of SMC with carbon fibers }
have been generated by the Leibniz-Institut für Verbundwerkstoffe as part of the project "C-SMC~Digitalization" funded by the Fraunhofer~ITWM within the framework of the High Performance Center Simulation and Software Based Innovation.

\bibliographystyle{unsrt}  
\bibliography{biblio,litbank-ml-2022}  

\section{Appendix}
\label{appendix:syntheticResults}

\end{document}